\documentclass[twocolumn,superscriptaddress,showpacs,aps,amsmath,amssymb]{revtex4}
\usepackage{epsfig}
\usepackage{amsbsy}
\usepackage{amsmath}
\usepackage{amsfonts}
\usepackage{graphicx}
\usepackage{latexsym}

\begin{document}

\title{High frequency limit for single-electron pumping operations}
\author{Chuan-Yu Lin}
\affiliation{Department of Physics and Center for Quantum
information Science, National Cheng Kung University, Tainan 70101,
Taiwan}
\author{Wei-Min Zhang}
\email{wzhang@mail.ncku.edu.tw} \affiliation{Department of Physics
and Center for Quantum information Science, National Cheng Kung
University, Tainan 70101, Taiwan}

\begin{abstract}
In this Letter, we study the transient electron transfer phenomena
of single-electron devices with alternating external gate voltages.
We obtain a high frequency limit for pumping electrons one at a time
in single-electron devices. Also, we find that in general the
electrical current is not proportional to the frequency of the
external signals in the single-electron devices, due to the strong
quantum coherence tunneling effect.
\end{abstract}

\date{Dec. 2,~2010}


 \maketitle

Single-electron pumps and turnstiles are nanoscale tunneling devices
utilizing controllable transfer of electrons one-by-one synchronized
with alternating external gate voltages. These devices are supposed
to have important applications as current standards and also as
high-frequency amplifiers/detectors in solid-state quantum
computing. Single-electron pumping operations have been
experimentally demonstrated with various nanoscale tunneling
structures \cite{set1,set2,set3,set4,set5,set6,Kem09172108,pumps}.
However, most of experimental realizations for single-electron
turnstiles are basically at the level of classical charge dynamics
with relatively low signal frequencies ($\sim$  tens to hundreds
MHz) and the relatively small pumped current ($\sim$ a few pA). To
achieve the device as a quantized source of electron current, we
shall closely monitor the electron transfer in single-electron
devices in the high frequency region to find the optimal conditions
for single-electron pumping operations.

Previous studies concerning electron transport in various
nanodevices have largely been focusing on the understanding and
prediction of the steady-state transport phenomena
\cite{Dat95,Jau08}. Time-dependent nonequilibrium transport are much
more complicated \cite{Jau945528,Mac06085324,Jin10083013}. However,
for a nanodevice, in particular, for quantum devices, the big
challenge is to understand and predict not only how fast or slow a
nanoscale device can turn on or off a current, but also how reliably
and efficiently the device can maintain the quantum coherence
through external field controls. Furthermore, it is also challenging
to manipulate the quantum device by driving the contacts far away
from equilibrium. These challenges require a full understanding of
the time-dependent quantum transport intimately entangling with
quantum coherence of electrons in the device. Therefore, in this
Letter, we shall utilize the nonequilibrium quantum transport theory
for nanodevices we developed recently \cite{Jin10083013} to monitor
the real-time dynamics of electronic transfer in single-electron
devices with high frequency external gate voltage, to understand the
features of devices far away from equilibrium.

\begin{figure}
\includegraphics[width=0.5\columnwidth,angle=0]{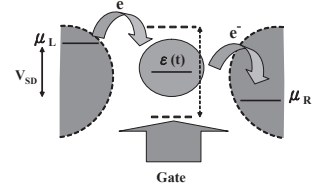}
\caption{A schematic plot of the single-electron turnstiles (a
quantum dot device coupled to two leads), and Gate control the
energy level of the dot in order to be the turnstile.} \label{fig1}
\end{figure}

A typical single-electron device consists of three serially
connected metallic islands and a gate coupled only to the central
island. Here we model the central island as a quantum dot with a
single energy level, as shown schematically by Fig.~\ref{fig1}. The
pump or turnstile operations are realized by imposing the repetitive
pulses to the gate. When an alternating voltage is applied to the
gate, electrons can be transferred one-by-one between the source and
the drain during every voltage pulse under certain conditions. We
will apply a harmonic time modulation to vary the energy level of
the dot,
$\varepsilon(t)=\varepsilon_0+\varepsilon_c\sin(\omega_ct)$, and a
dc bias voltage $V_{\rm SD}$ between the source and drain to examine
the electron transfer dynamics.

Based on the recently developed non-equilibrium quantum transport
theory for nanodevices \cite{Jin10083013}, the electron occupation
number in the dot and the transient electron current from the leads
into the dot are given by
\begin{subequations}
\label{curr-char}
\begin{align}
 n(t) = &~ v(t,t)+ u(t,t_0)n(t_0)u^{\dag}(t,t_0), \\
 I_{L,R}(t)= & -{2e\over \hbar}{\rm Re}\int_{t_0}^{t} d\tau {\rm
Tr}\Big\{g_{L,R}(t,\tau)v(\tau,t)-\widetilde{g}_{L,R}(t,\tau)\notag
\\ & \times u^\dag(t,\tau)
+g_{L,R}(t,\tau)u(\tau,t_0)n(t_0)u^{\dag}(t, t_0)\Big\},
\end{align}
\end{subequations}
respectively, where $L,R$ denote the left and right leads (source
and drain).  The functions $u(\tau, t_0)$ and $v(\tau, t)$ in
Eq.~(\ref{curr-char}) are related to the retarded and correlation
Green functions that satisfy the integrodifferential equations of
motion \cite{Tu08235311,Jin10083013}:
\begin{subequations}
\label{uv-eq}
\begin{align}
&\dot{u}(\tau, t_0) +i \varepsilon(\tau) u (\tau,
t_0)+\int_{t_0}^{\tau } d\tau' g
(\tau ,\tau') u (\tau', t_0) =0 ,  \label{u-e} \\
&v(\tau, t)=\int_{t_0}^{\tau} d\tau' \int_{t_0}^{t} d\tau'' u (\tau,
\tau') \widetilde{g}(\tau', \tau'') u^{\dag}(t,\tau'') , \label{v-e}
\end{align}
\end{subequations}
subjected to the initial condition $u(t_0,t_0)=1$. $n(t_0)$ in
Eq.~(\ref{curr-char}) is the initial electron occupation in the dot.
Here, we have defined $g (\tau ,\tau')=\sum_{\alpha=L,R} g_\alpha
(\tau ,\tau')$ and $ \widetilde{g} (\tau ,\tau')=\sum_{\alpha=L,R}
\widetilde{g}_\alpha (\tau ,\tau')$. $g_{L,R}( \tau ,\tau')$ and
$\widetilde{g}_{L,R}(\tau ,\tau')$ are the time-correlation of
electron transferring in the leads
 through the dot \cite{Tu08235311}:
\begin{subequations}
\label{correlation}
\begin{align}
g_{L,R}( \tau ,\tau') &=\int_{-\infty}^{\infty }\frac{d\omega}{2\pi}
J_{L,R}\left(
\omega \right) e^{-i\omega ( \tau -\tau')} ,   \\
\widetilde{g}_{L,R}(\tau ,\tau') &=\int_{-\infty}^{\infty
}\frac{d\omega}{2\pi} J_{L,R}( \omega) f_{L,R}(\omega)e^{-i\omega (
\tau -\tau')} ,
\end{align}
\end{subequations}
in which $f_{L,R}(\omega) =\frac{1}{e^{\beta (\omega-\mu_{L,R})}
+1}$ are the initial electron distribution functions in the leads at
the initial temperature $\beta=1/k_BT$, and  $\mu_{L,R}$ the
corresponding chemical potentials. $J_{L,R}(\omega)=2\pi
\rho_{L,R}(\omega)|V_{L,R}(\omega)|^2$ are the spectral densities
with $\rho_{L,R}(\omega)$ being the densities of states of the leads
$L$ and $R$, and $V_{L,R}(\omega)$ the lead-dot coupling
coefficients. In reality, most of spectral densities have more or
less a Lorentzian-type shape,
\begin{align}
J_{L,R}(\omega) = \frac{\Gamma_{L,R} d_{L,R}^{2}}{(\omega
-\mu_{L,R})^{2}+d_{L,R}^{2}},\label{spectral}
\end{align}
where $\Gamma_{L,R}$ are the electron tunneling rate from the leads
to the dot, and $d_{L,R}$ are the bandwidths of the spectral
densities. The integral kernels, $g_{L,R}( \tau ,\tau')$ and
$\widetilde{g}_{L,R}(\tau ,\tau')$,  characterize all the back
action memory effects between the leads and dot associating with
quantum dissipation and fluctuation. These effects must be fully
taken into account for the accuracy of single-electron transfer.

\begin{figure}
\includegraphics[width=0.75\columnwidth,angle=0]{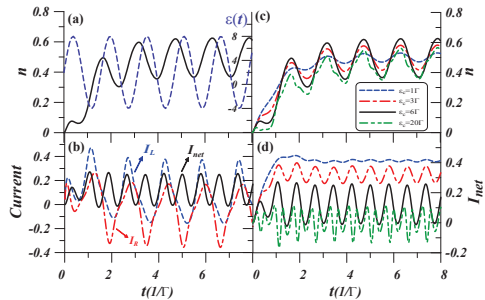}\newline
\caption{The time-dependent electron transfer phenomena in the
single-electron turnstile. (a) The Harmonic time modulation (the
dotted line): $\varepsilon(t)=\varepsilon_0+\varepsilon_c
\sin(\omega_ct)$ with $\varepsilon_0=2\Gamma$,
$\varepsilon_c=6\Gamma$ and $\omega_c=4 \Gamma$, and the electron
population (the solid line) in the dot; (b) the corresponding
electron transfer currents $I_L(t)$, $I_R(t)$ and $I_{\rm net}(t)$;
(c) and (d) are the electron population and the net current with
different signal strength $\varepsilon_c$, respectively.}
\label{fig2}
\end{figure}

\begin{figure}
\includegraphics[width=0.75\columnwidth,angle=0]{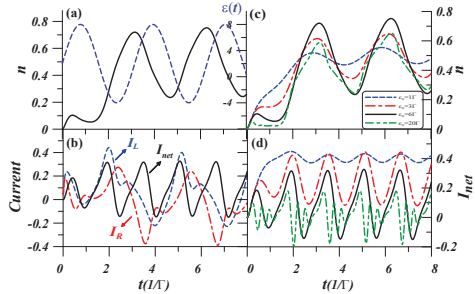}\newline
\caption{The same plots as in Fig.~\ref{fig2} with a different
signal frequency $\omega_c=2\Gamma$.} \label{fig3}
\end{figure}

\begin{figure}
\includegraphics[width=0.75\columnwidth,angle=0]{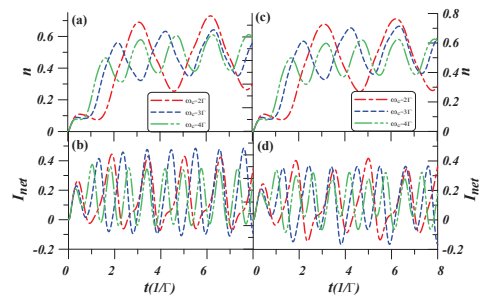}
\caption{Electron population and the net current for different bias
voltages. (a)-(b) The bias $eV_{\rm SD}=6 \Gamma$ is symmetrically
applied to the leads with $\varepsilon_0=3\Gamma$. (c)-(d) The bias
$eV_{\rm SD}=5 \Gamma$ is asymmetrically applied to the leads with
$\varepsilon_0=2\Gamma$. Here $\varepsilon_c=6 \Gamma$} \label{fig4}
\end{figure}

\begin{figure}
\includegraphics[width=0.65\columnwidth,angle=0]{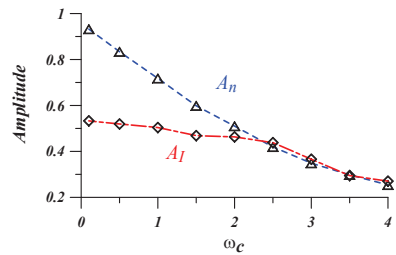}
\caption{The oscillation amplitudes of the electron population in
the dot (given by $A_n$) and the current flowing from the left to
right leads (given by $A_I$) with varying the  signal frequency
$\omega_c$. Here $eV_{\rm SD}=4\Gamma$, $\varepsilon_0=2\Gamma$ and
$\varepsilon_c=6\Gamma$. } \label{fig5}
\end{figure}

The above transient quantum transport theory can reproduce the
time-dependent transport theory of mesoscopic systems developed by
Jauho {\it et al.} based on Keldysh's nonequilibrium Green function
technique \cite{Jau945528}. The main advantage of the present theory
is that it takes into account explicitly the initial state
dependence of the device \cite{Jin10083013}. Thus the time-dependent
electron transfer in single-electron devices can be monitored with
alternating external gate voltages from an arbitrary initial state
of the device. In the following calculation, we take $n(t_0)=0$. The
detailed results are plotted in Figs.~\ref{fig2}-\ref{fig5}, where
we take the tunneling rate $\Gamma_L=\Gamma_R=\frac{1}{2}\Gamma$.
The applied bias voltage $V_{\rm SD}$ is set to be a constant, and
the dot energy level varies with time:
$\varepsilon(t)=\varepsilon_0+\varepsilon_c \sin(\omega_ct)$, where
the  signal frequency will be set as a function of $\Gamma$ to
examine the validity of the high frequency operation for pumps and
turnstiles. The initial temperature of the leads is taken at
$k_BT=0.1\Gamma$. We also fix the bandwidth of the spectral density:
$d_{L,R}=20 \Gamma$ which is close to the wide band limit. All the
parameters can be controlled experimentally. In the rest of the
Letter, we will focus on the electron transfer phenomena of the
device at different input parameters to understand the transient
electron dynamics in the single-electron devices.

Fig.~\ref{fig2} plots the electron population in the dot, the left
and right current flowing into the dot as well as the net current
passing through the dot at the signal frequency $\omega_c=4\Gamma$
with different signal strength (amplitude) $\varepsilon_c$. The bias
voltage $eV_{SD}=4 \Gamma$ is added symmetrically to the source and
drain while the energy level of the dot is set
$\varepsilon_0=2\Gamma$. In Fig.~\ref{fig2}(a), the dotted line is
plotted for the time dependence of the  signal (harmonic modulation)
and the solid line is for the electron population in the dot. The
result shows that the electron population in the dot oscillates
between $0.3 \sim 0.7$ with the same frequency as that of the
signal, except for a phase shift. The electron population does not
vary between zero to one is due to the quantum coherent tunneling of
the electron between the leads and the dot. The currents are plotted
in Fig.~\ref{fig2}(b). The results show that the currents $I_L(t)$
and $I_R(t)$ oscillate with the same signal frequency but the
oscillating shapes are deformed slightly from the sinusoid with a
rather small phase shift. The net current $I(t)$ which depicts the
electron transfer from the left to right leads shows to be a perfect
sinusoidal oscillation, with the oscillating frequency being the
twice of the signal frequency, as a result of symmetrically applying
the harmonic modulation to the dot with respect to the source and
drain. In Fig.~\ref{fig2}(c) we plot the time dependence of the
electron population at different signal strengthes. As one can see,
increasing the signal amplitude $\varepsilon_c$ enhances the
oscillating amplitude of the electron population. It is interesting
to find that when $\varepsilon_c \simeq 6\Gamma$, the oscillating
amplitude of the electron population reaches the maximal value.
Continuously increasing the signal strength will weaken the
population oscillations. The same situation happens for the current,
the current oscillation amplitude reaches the maximal value at
$\varepsilon_c \simeq 6\Gamma$. Further increasing the signal
strength will decrease the strength of the current oscillation.

Next, we shall vary the signal frequency to see the corresponding
change of the transfer phenomena. Fig. 3 plots the same physical
quantities at the signal frequency $\omega_c=2\Gamma$. The electron
population in the dot behaves almost the same, see
Fig.~\ref{fig3}(a).  The left and right currents $I_L(t)$ and
$I_R(t)$ still oscillate with the signal frequency but a
photon-assisted peak appears as a nonlinear response to the signal.
The net current $I(t)$ remains in a good oscillating profile with
the oscillating frequency still being the twice of the signal
frequency, as shown in Fig.~\ref{fig3}(b). Fig.~\ref{fig3}(c)-(d)
plot the electron population and the net current with different
signal strengthes. Again, we find that the signal strength for
maximal amplifying locates at $\varepsilon_c \simeq 6\Gamma$ where
both the electron population and current oscillate with maximal
amplitudes. Continuously increasing the signal strength only induces
more photon-assisted peaks but does not amplify the transfer
current, where the photon-assisted peaks occur near $\varepsilon =
\varepsilon_0 \pm 2k\omega_c$ and $k$ is an integer.

To further analyze the controllable electron transfer in the device,
we vary the dc bias voltage $V_{\rm SD}$ applied to the leads.
Fig.~\ref{fig4} plots the electron population and the net current
with different bias. In Fig.~(\ref{fig4}(a)-(b), the bias $eV_{\rm
SD}=6 \Gamma$ is symmetrically applied to the leads with
$\varepsilon_0=3\Gamma$, while in Fig.~(\ref{fig4}(c)-(d), the bias
$eV_{\rm SD}=5 \Gamma$ is asymmetrically applied to the leads with
$\varepsilon_0=2\Gamma$. As we see the electron population and the
net current do not show a qualitative difference in both cases.
However, the asymmetric case do change the profile of the net
current due to the breaking of the symmetry.

Comparison with the results in Figs.~\ref{fig2}-\ref{fig4}, we find
that amplifying the electron transfer in the single-electron devices
is mainly controlled by the signal frequency and the signal
strength. The signal strength $\varepsilon_c \simeq 6\Gamma$ is an
optimal operation condition for the single-electron device reaching
the maximal amplifying. In Fig.~\ref{fig5}, we plot the maximal
oscillating amplitudes of the electron population and the net
current varying with different signal frequency $\omega_c$ at
$\varepsilon_c = 6\Gamma$. As we see the oscillating amplitude of
electron population increases with the decreasing of the signal
frequency $\omega_c$. It will approach to almost one with a relative
low frequency ($\omega_c < 0.1 \Gamma$). This provides a high signal
frequency limit (in terms of tunneling rate $\Gamma$) for the single
electron device acting as a single-electron pump. With higher
frequencies, can an electron only be partially occupied in the dot,
due to the strong quantum coherence of electrons between the leads
and dot. In other words, the electron occupation number in the dot
can never be perfectly one or zero in the high frequency region
($\sim \Gamma)$. The oscillating amplitude of the net current also
increases with decreasing the signal frequency $\omega_c$ but when
$\omega_c < 2\Gamma$, the oscillating amplitude of the net current
approaches to a constant, as shown in Fig.~\ref{fig5}. This
indicates that the electrical current is not proportional to the
frequency of the external signals. Taking the experimental value
$\Gamma=50 \mu eV$ in \cite{set4}, the high frequency limit for
single-electron pumps is $\omega_c \simeq 7.5$MHz. For the
experimental data $\Gamma=50 meV$ \cite{set5}, single-electron pumps
works for the signal frequency $\omega_c <0.1 \Gamma \simeq 7.5$GHz.
These solutions are consistent with the current experimental results
\cite{set4,set5}.

In conclusion, the above analysis and discussion show a general
picture how the electron population in the dot and the current
passing through the dot response to the frequency and the strength
of the alternating external gate voltage. With a harmonic time
modulation, we find that the optimal operations for the single
electron device is near the signal strength $\varepsilon_c \simeq
6\Gamma$. Due to the strong coherence tunneling at high frequency
regime, the single-electron pumps or turnstiles works only at the
signal frequency $\omega_c<0.1 \Gamma$. Realistic single-electron
devices involve much more complicated nanostructures
\cite{set1,set2,set3,set4,set5,set6,Kem09172108,pumps} than the
simplified model considered in the present work, but our rigorous
analysis could provide a useful guide for high frequency operations
of single-electron pumps and turnstiles. On the other hand, the
strong quantum coherence of the electron transfer in high frequency
region shows that single-electron device operations at high
frequency are promising for quantum information processing
\cite{QIP}.

This work is supported by the National Science Council of ROC under
Contract No. NSC-99-2112-M-006-008-MY3 and National Center for
Theoretical Science.


\begin{thebibliography}{9}

\bibitem{set1} L. J. Geerligs, V. F. Anderegg, P. A. M. Holweg, J. E. Mooij, H.
Pothier, D. Esteve, C. Urbina, and M. H. Devoret, Phys. Rev. Lett.
\textbf{64}, 2691 (1990).

\bibitem{set2} M. W. Keller, A. L. Eichenberger, J. M. Martinis, and N. M.
Zimmerman, Science \textbf{285}, 1706 (1999).

\bibitem{set3} S. V. Lotkhov, S. A. Bogoslovsky, A. B. Zorin, and J. Niemeyer,
Appl. Phys. Lett. \textbf{78}, 946 (2001).

\bibitem{set4} A. Fujiwara, N. M. Zimmerman, Y. Ono, and Y. Takahashi, Appl.
Phys. Lett. \textbf{84}, 1323 (2004); A. Fujiwara, K. Nishiguchi and
Y. Ono, Appl. Phys. Lett. \textbf{92} 042102 (2008).

\bibitem{set5} M. D. Blumenthal, B. Kaestner, L. Li, S. Giblin, T. J. B. M. Janssen,
M. Pepper, D. Anderson, G. Jones and D. A. Ritchie, Nat. Phys.
\textbf{3}, 343 (2007).

\bibitem{set6} J. P. Pekola, J. J. Vartiainen, M. M\"{o}tt\"{o}nen, O.-P. Saira, M.
Meschke, and D. V. Averin, Nat. Phys. \textbf{4}, 120 (2008).

\bibitem{Kem09172108}A. Kemppinen, S. Kafanov, Yu A. Pashkin, J. S. Tsai, D. V. Averin
and J. P. Pekola, Appl. Phys. Lett. \textbf{94} 172108 (2009).

\bibitem{pumps}S. P. Giblin, S. J. Wright, J. D. Fletcher, M. Kataoka, M. Pepper,
T. J. B. M. Janssen, D. A. Ritchie, C. A. Nicoll, D. Anderson, and
G. A. C. Jones, New J. Phys. \textbf{12}, 073013 (2010).

\bibitem{Dat95} S. Datta, \textit{Electronic Transport in Mesoscopic Systems},
(Cambridge, England, 1995).

\bibitem{Jau08} H. Haug and A. P. Jauho, \textit{Quantum Kinetics in
Transport and Optics of Semiconductors}, Springer Series in
Solid-State Sciences 123, 2nd Ed. (Springer-Verlag, Berlin, 2008).

\bibitem{Jau945528} A.-P. Jauho, N. S. Wingreen, and Y. Meir, Phys. Rev. B
\textbf{50}, 5528 (1994).

\bibitem{Mac06085324} J. Maciejko, J. Wang, and H.
Guo, Phys. Rev. B \textbf{74}, 085324 (2006).


\bibitem{Jin10083013} J. S. Jin, M. W. Y. Tu, W. M. Zhang,
Y. J. Yan, New J. Phys. \textbf{12}, 083013 (2010).

\bibitem{Tu08235311} M. W. Y. Tu and W. M. Zhang, Phys. Rev. B \textbf{78}, 235311 (2008).


\bibitem{QIP} G. Feve, A. Mahe, J.-M. Berroir, T. Kontos, B. Placais,
D. C. Glattli, A. Cavanna, B. Etienne and Y. Jin, Science
\textbf{316}, 1169 (2007).








\end{thebibliography}
\end{document}